\documentclass[%
reprint,
showkeys,
amsmath,amssymb,
aps,
pra,superscriptaddress
]{revtex4-2}

\usepackage{graphics}
\usepackage{epsfig}
\usepackage{epstopdf}
\usepackage{amsmath}
\usepackage{array}
\usepackage{subfigure}
\usepackage[colorlinks=true,linkcolor=blue,citecolor=blue,      urlcolor=blue]{hyperref}
\usepackage{siunitx,scalerel,float}
\begin{document}
	\sloppy	
	\title{Thermal evolution and stability analysis of phenomenologically emergent dark energy model
	}
	
	\title{Thermal evolution and stability analysis of phenomenologically emergent dark energy model}        
	
\author{Rosemin John}
\email{roseminjohn13@gmail.com}
\affiliation{
	Department of Physics, Cochin University of Science and Technology, Kochi, Kerala 682022, India
}
\author{Sarath N.}
\email{sarathn@iitk.ac.in}
\affiliation{
	Indian Institute of Technology, Kanpur 208016, India
}
\author{Titus K. Mathew.}
\email{titus@cusat.ac.in}
\affiliation{
	Department of Physics, Cochin University of Science and Technology, Kochi, Kerala 682022, India
}

			\begin{center}
		\begin{abstract}
			The phenomenologically emergent dark energy (PEDE) model is a varying dark energy model with no extra degrees of freedom proposed by Li and Shafieloo\citep{Li_2019} to alleviate the Hubble tension. The statistical consistency of the model has been discussed by many authors. Since the model depicts a phantom dark energy that increases with redshift, its cosmic evolution, particularly during the late phase, must be examined. We discover that the model's Hubble and deceleration parameters display unusual behaviour in the future, which differs from $\Lambda$CDM cosmology. We find the model also follows a distinct evolution in the statefinder plane. The phantom nature of the model leads to the violation of the null energy condition and a decrease in horizon entropy. The asymptotic future epoch also seems to be unstable based on our dynamical system analysis as well as the stability analysis based on dark energy sound speed.		
			\keywords{PEDE model -- NEC -- GSL -- entropy -- stability}
		\end{abstract}
	\maketitle
		\end{center}
		\section{Introduction}
		
		The late acceleration of the universe's expansion was one of the crucial cosmological discoveries \citep{riess1998observational,perlmutter1999measurements}. This phenomenon couldn't be explained if the universe is composed only of matter and radiation, which prompted scientists to look for a new exotic cosmic component with a negative equation of state, $\omega$ ($\omega=p/\rho$, $p$ is the pressure and $\rho$ is the density), that they termed `dark energy'. The current standard model, the $\Lambda$CDM model, incorporates cosmological constant as dark energy, although its precise nature has not yet been established. Despite being able to explain the late acceleration in the universe's expansion, the $\Lambda$CDM model has some inconsistencies \citep{Perivolaropoulos_2022, bull2016beyond}. The cosmological constant problem \citep{weinberg1989cosmological}, also known as the fine tuning problem, stems from the enormous gap between the dark energy density that was determined from quantum field theory and that obtained from cosmic observations. The difference is around $10^{121}$ orders of magnitude \citep{copeland2006dynamics}. The coincidence problem \citep{sivanandam2013cosmological, velten2014aspects}, which concerns the identical densities of matter and dark energy in the current universe although these components evolve differently such that the matter density decreases in inverse proportion to the volume of the universe while the density of the cosmological constant remains constant throughout evolution, is another significant issue with $\Lambda$CDM cosmology. There also exists tensions in the parameter values extracted from cosmic data, among which the recently discovered Hubble tension has gained the most attention.
		`Hubble tension' mainly refers to the difference in the present value of the Hubble parameter extracted from cosmic microwave background (CMB) data and that from local measurements. The Planck collaboration \citep{aghanim2020planck}, from the CMB observations assuming $\Lambda$CDM, provided a value of  $H_{0}=67.4 \pm 0.5\ \si{ kms^{-1}Mpc^{-1}}$.
		The SH0ES (Supernova H0 for the Equation of State) group of Hubble space telescope, from the observations of 70 long-period Cepheids in the Large Magellanic Cloud with other local distance anchors predicted the value of  $H_{0}=74.03 \pm 1.42\ \si{ kms^{-1}Mpc^{-1}}$  \citep{Riess_2019}. The inconsistency between these two measurements is over 4$\sigma$. The tension has now increased over time as a result of examining multiple data sets gathered using various techniques
		\citep{Di_Valentino_2021, odintsov2021analyzing,  Perivolaropoulos_2022}. If there are no systematic errors and since the CMB measurements are model dependent, it is anticipated that the existing standard cosmological model will be required modifications to account for this tension.  
		
		There have been several attempts to alleviate these problems of the $\Lambda$CDM model. The easiest and most natural way is to assume a different form for the dark energy.
		Such dark energy models include: scalar field k-essence models \citep{PhysRevLett.85.4438}, quintessence dark energy models with scalar field having $\omega>-1$ \citep{tsujikawa2013quintessence, dedeo2003effects}, phantom models where dark energy is replaced with a phantom component which has $\omega<-1$ \citep{Caldwell_2002, caldwell2003phantom},  quintom model of dark energy with two scalar fields with one being quintessence and the other phantom \citep{Guo_2005, cai2010quintom}, interacting dark energy models \citep{PhysRevD.94.123511, PhysRevD.101.103533, Di_Valentino_2020}, Chaplygin gas model \citep{chaplygin,  Amendola_2003, PhysRevD.69.123524}, 
		holographic dark energy model \citep{li2004model}, etc.
		Modified gravity theories\citep{RevModPhys.82.451, De_Felice_2010, Nojiri_2017, nunes2018structure,  rossi2019cosmological, braglia2020larger, Ballardini_2020, adi2021can, Yan_2020} can be another significant approach to address the challenges faced by the standard model.
		
		Recently, Li and Shafieloo \citep{Li_2019} proposed a new dark energy model called the Phenomenologically Emergent Dark Energy (PEDE) model, in which dark energy has a varying density and a phantom nature. It was primarily developed to address the Hubble tension. The model effectively replaces the cosmological constant with a hyperbolic tangent function of redshift which causes the dark energy to emerge as a function of the cosmic time at later times. The absence of additional parameters makes the model more intriguing. PEDE model using the Planck 2018 CMB data + Pantheon + BAO + Ly-$\alpha$ data finds $H_{0}=71.02_{-1.37}^{+1.45}\ \si{ kms^{-1}Mpc^{-1}}$  \citep{Li_2019} thereby resolving the tension on $H_0$ within 68\% confidence-level.
		The higher Hubble constant value is a direct consequence of dominance of the phantom energy during the late phase. Later, a generalized PEDE model with an extra degree of freedom was proposed \citep{Li_2020}, and under the right limiting conditions, it can be reduced to both the original PEDE model and the $\Lambda$CDM model. Benaoum et al.\cite{benaoum2022modified} have also developed a modified version of the PEDE model known as Modified Emergent Dark Energy (MEDE), which also claims to incorporate the $\Lambda$CDM, PEDE model, Chevallier-Polarski-Linder (CPL) model and other related cosmological models.
		
		The PEDE model was further analyzed statistically in comparison with the cosmological data by various authors. Yang et al.\cite{yang2021emergent} claim that the model can still reduce cosmic tensions despite the inclusion of massive neutrino species and possible additional relativistic degrees of freedom in their revised analysis of the PEDE scenario. The Hubble constant tension is found to be below the 2$\sigma$ level in this scenario, and the value of $\sigma_8$ is in better agreement with cosmic shear estimates. They have obtained $M_{\nu}\sim 0.21_{-0.14}^{+0.15}$ eV and $N_{eff}=3.03\pm 0.32$,  which point to a non-zero neutrino mass with a significance greater than 2$\sigma$. However, they obtained the sound horizon as $\sim$ 147 Mpc, failing to restore the lower value that the BAO data prefers.	 
		In Ref.\cite{Rezaei_2020}, the authors have demonstrated how the PEDE model can solve the coincidence problem along with Hubble tension. But the $\sigma_8$ tension between low-redshift observations and Planck inferred value, however, cannot be resolved by the PEDE model, and the analysis with the growth rate dataset leads them to the conclusion that the PEDE model is less capable of explaining observable evidence in cluster scales at the perturbation level when compared to the $\Lambda$CDM.
		According to the Bayesian analysis performed on the model by Pan et al.\cite{pan2020reconciling}, the majority of observational dataset combinations favor $\Lambda$CDM over the PEDE model.
		Alestas et al.\cite{PhysRevD.105.063538} have also analyzed the model's viability through the Akaike information criterion (AIC) and the Bayes factor considering the full cosmic microwave background temperature anisotropy spectrum data with other cosmological datasets and they found the model is strongly disfavored in both criteria. 	 
		Sch\"{o}neberg et al.\cite{olympics} have shown PEDE model offers a worse fit to the combined datasets (Planck 2018 (including TTTEEE and lensing) + BAO (including BOSS DR12 + MGS +6dFGS)+Pantheon) than $\Lambda$CDM. This is due to a significant degradation of the joint fit to BAO and Pantheon data and is a result of the non-nested nature of the PEDE model. The model also decreases the angular diameter distance for the CMB, because of the equation of state $\omega<-1$ that this model exhibits at later times. The authors also point out that the model has the ability to artificially lower the tension while severely impairing the fit to the CMB anisotropy data.
		Sharma\cite{SHARMA2020100717} studied the PEDE model and its anisotropic extension  ($\mathcal{A}$PEDE) via the Bianchi type-I spacetime metric. When compared to the base $\Lambda$CDM, the statistical results for both the PEDE and $\mathcal{A}$PEDE models indicate a good fit with the Hubble+CMB data combination, however, the addition of BAO or Pantheon with other data combinations lead to very strong evidence against the PEDE models.
		
		Even while the PEDE model's cosmological evolution has been extensively studied, its thermodynamic evolution has not been considered so far. Therefore, the primary purpose of this research is to investigate the thermal evolution of the PEDE model. We further perform a stability analysis of this model intending to compare its feasibility in comparison with the standard $\Lambda$CDM. In the following part, we briefly present the model before extracting the parameter values. We begin our analysis by obtaining an analytical solution for the Hubble and deceleration parameters of the PEDE model and study their evolution in detail. We then use the statefinder diagnostic to distinguish the PEDE model from other cosmological models. We then continue to analyze the status of the laws of thermodynamics. We examine the Generalized Second Law(GSL) of thermodynamics and the convexity condition to check whether the model implies maximization of the horizon entropy. Finally, employing dynamical system analysis and sound speed analysis, we investigate the stability of the model.
		\section{Phenomenologically Emergent Dark Energy}
		Within the framework of Einstein's general relativity, a spatially flat, homogeneous and isotropic universe following the Friedmann-Lema\^itre-Robertson-Walker (FLRW) metric can be explained by the Friedmann equation,
		\begin{equation}
			\label{F1}
			H^2 =\frac{8\pi G}{3}(\rho_\gamma+\rho_m + \rho_{\scaleto{DE}{3.5pt}})
		\end{equation}
		where $H$ is the Hubble parameter and $G$ is the gravitational constant. $\rho_\gamma,\rho_m$ and $\rho_{\scaleto{DE}{3.5pt}}$ represent the density 
		of radiation, matter and dark energy respectively. The density components can be expressed in terms of their corresponding dimensionless density parameters, $\Omega_i=\rho_i/\rho_{c,0}$ where $\rho_{c,0}=3H_0^2/8\pi G$ is the critical density. Eq.\ref{F1} can then be rewritten as,
		\begin{equation} \label{H} 	
			H^2(z)= H_0^2\left[\Omega_{\gamma0}(1+z)^4+\Omega_{m0}(1+z)^3+\Omega_{\scaleto{DE}{3.5pt}}(z)\right].
		\end{equation}
		Naught in the subscript denotes the present value of the parameters. In the standard model, the dark energy density parameter has a constant value, $\Omega_{\scaleto{DE}{3.5pt},0}$. But, in general the evolution of  $\Omega_{\scaleto{DE}{3.5pt}}(z)$ with redshift can be expressed as,
		\begin{equation}  	
			\Omega_{\scaleto{DE}{3.5pt}}(z)=\Omega_{\scaleto{DE}{3.5pt},0}\times\exp\Bigg[3\int_{0}^{z}\frac{1+\omega_{\scaleto{DE}{3.5pt}}(z')}{1+z'}dz'\Bigg].
		\end{equation}
		where $\omega_{\scaleto{DE}{3.5pt}}$ is the equation of state of dark energy. Therefore, one can construct a dark energy model by assuming a specific form for either equation of state or density of dark energy. In PEDE model, Li and Shafieloo\cite{Li_2019} have given the following form for dark energy density:
		\begin{equation}  	
			\label{pede de}
			\Omega_{\scaleto{DE}{3.5pt}}(z)=\Omega_{\scaleto{DE}{3.5pt},0}\times\Big(1-\tanh\big[\log_{10}(1+z)\big]\Big).
		\end{equation}
		As the authors claim, in this model, dark energy has no effective presence in the past as its density will be negligibly small, but emerges at later times as the universe evolves. Therefore, in this study, we are mainly interested in the PEDE model's late phase evolution, during which dark energy becomes the dominant component. (We disregard the contribution of radiation from here onward because $\rho_ \gamma$ will be negligibly small in late phase.)
		
		If matter and dark energy have no interaction, then the conservation equations are as follows:
		\begin{equation}  	
			\begin{split}\label{cont}
				\Dot{\rho}_{m}+3H\rho_{m}(1+\omega_{m})=0 \\
				\Dot{\rho}_{\scaleto{DE}{3.5pt}}+3H\rho_{\scaleto{DE}{3.5pt}}(1+\omega_{\scaleto{DE}{3.5pt}})=0 
			\end{split}
		\end{equation}	
		where $\omega_m = 0$ and $\omega_{\scaleto{DE}{3.5pt}}$ can be derived as,
		\begin{equation}  	
			\omega_{\scaleto{DE}{3.5pt}}(z)=\frac{1}{3}\frac{d \ln \Omega_{\scaleto{DE}{3.5pt}}}{dz}(1+z)-1.
		\end{equation}
		Substituting the PE dark energy and the equation becomes, 
		\begin{equation}
			\omega_{\scaleto{DE}{3.5pt}}(z)	=-\frac{1}{3\ln10}\times\Big(1+\tanh\big[\log_{10}(1+z)\big]\Big)-1
		\end{equation}
		Both $\Omega_{\scaleto{DE}{3.5pt}}$ and $\omega_{\scaleto{DE}{3.5pt}}$ follows a symmetrical behavior in logarithmic scale(Fig.1 in \citep{Li_2019}).
		For an early epoch, i.e., when $z$ is large and positive, $\omega_{\scaleto{DE}{3pt}}$ converges to a value of $-\frac{2}{3\ln10}-1$. As time progresses the equation of state increases, becomes  $-\frac{1}{3\ln10}-1$ in the present epoch, and asymptotically reaches $-1$ in the far future. Here the interesting fact to be noted is that the PEDE is having phantom nature with $\omega_{\scaleto{DE}{3pt}}<-1$\citep{CALDWELL200223} except at the asymptotic limit, $z \to -1.$ 
		This is in clear contrast with the standard model, for which the equation of state of dark energy stays pegged at -1.
		
		\section{Parameter Estimation}
		We have estimated and compared the standard $\Lambda$CDM and  PEDE model parameters using the cosmological observation data on : $H(z)$ data \citep{Geng_2018}, Pantheon sample consisting of type Ia supernovae data \citep{scolnic2018complete}, cosmic microwave background (CMB) data \citep{chen2019distance} and baryon acoustic oscillation (BAO) data \citep{10.1111/j.1365-2966.2011.19592.x}. The parameters are obtained by applying Markov Chain Monte Carlo (MCMC) method using the emcee package as optimizer \citep{Foreman-Mackey_2013} in the lmfit python library \citep{matt_newville_2021_4516644}.  We have adopted uniform priors for the parameters as, $H_0$ in the range, 65 to 75 $ \si{ kms^{-1}Mpc^{-1}}$ and $\Omega_{m0}$ in the range, 0.1 to 0.5 .
		The best fit values are then calculated by minimizing the $\chi^2$ function defined as \citep{N_2021,Dheepika_2022},
		\begin{equation}
			\chi^2_i = \sum\Big[\frac{V_i^{obs} - V_i^{th}}{\sigma_i}\Big]^2
		\end{equation}
		where $V_i^{obs}$ is the observed value from the datasets, $V_i^{th}$ is the corresponding
		theoretical value obtained from our model and $\sigma_i$ is the error in the measured
		values. The minimum $\chi^2$ per degrees of freedom or the reduced $\chi^2$ is defined as, $\chi^2_{d.o.f} = \frac{\chi^2}{N-N_p}$, where $N$ is the number of data points and $N_p$ is the number of variable parameters. We have also included the Akaike Information Criterion (AIC) and Bayesian Information Criterion (BIC), which are defined as,
		\begin{equation}
			\begin{split}
				AIC &= N \ln (\chi^2/N) + 2N_p\\
				BIC &= N \ln (\chi^2/N) + \ln(N)N_p
			\end{split}
		\end{equation}
		Similar to $\chi^2$, models with smaller AIC and BIC values are preferred. The observations considered for the estimation are detailed below.
		\begin{itemize}
			\item \textbf {$H(z)$ data:} The Hubble data set used in this analysis consists of 52 $H(z)$ measurements in the range $0\leq z\leq 2.36$ out of which 31 data points are from the differential age (DA) and 20 data points from clustering measurements\citep{Geng_2018}. The DA method, introduced by  Jimenez and Loeb\cite{Jimenez_2002}, measures the Hubble parameter by comparing the ages of passively-evolving galaxies with similar metallicity and separated by a small redshift interval. The latter is the clustering of galaxies or quasars which was first proposed in Ref.\citep{Gazta_aga_2009}, using the BAO peak position as a standard ruler in the radial direction. 
			The remaining data point is the direct measurement result of Hubble constant, $H_0=73.24\pm1.74$  \si{ kms^{-1}Mpc^{-1}}\citep{Riess_2016} from the HST. 
			\item \textbf {SN Ia:} The type Ia supernovae dataset used for the analysis is the latest Pantheon sample \citep{scolnic2018complete}  which comprises the luminosity of 1048 SNe Ia within the redshift $0.01<z<2.3$.  
			To calculate the observed distance modulus, Scolnic et al.\cite{scolnic2018complete} used the SALT2\citep{Guy_2010} light curve fitter,
			\begin{equation}
				\mu_i^{obs} = m(z)  + \alpha  X_1 - \beta  C - M
				\label{obs}
			\end{equation}
			where $m$ is the rest frame B-band peak magnitude and $M$ is the absolute B-band magnitude of a fiducial SN Ia with $X_1$ (time stretch of light curve) and $C$ (supernova color at maximum brightness) set to zero. The stretch-luminosity parameter ($\alpha$) and the
			color-luminosity parameter ($\beta$) are calibrated to zero for the pantheon sample, hence the observed distance modulus reduces to $\mu_i^{obs} = m  - M$. For our analysis, we treat $M$ as a nuisance parameter and is allowed to vary in the range, $-20\leq M \leq -18$.
			
			The luminosity distance of these SN Ia can be calculated from the equation,
			\begin{equation}
				d_L(H_0,\Omega_{m0},z_i) = c(1+z_i) \int_{0}^{z_i}\frac{dz'}{H(H_0,\Omega_{m0},z')} 
			\end{equation}
			where $z_i$ is the redshift of the SN Ia, $c$ is the speed of light and $H(H_0,\Omega_{m0},z')$ is the Hubble parameter specified for the model. The theoretical distance modulus of SN Ia is then given by,
			\begin{equation}
				\mu_i^{th}(H_0,\Omega_{m0},z_i) = 5\log_{10}\Big[\frac{d_L(H_0,\Omega_{m0},z_i)}{Mpc}\Big]+25.
			\end{equation}
			\item \textbf {CMB:} The theoretical shift parameter $R$ of the cosmic microwave background is defined as,
			\begin{equation}
				R(H_0,\Omega_{m0}) = \sqrt{\Omega_{m0}}\int_{0}^{z_*}\frac{dz}{h(H_0,\Omega_{m0},z)}
			\end{equation} 
			where $z_{*}$ is the redshift at the photon decoupling epoch and $h=H/H_0$. The Planck 2018 \citep{chen2019distance} observations suggests the distance prior measurement value $R^{obs}=1.7502\pm0.0046$ at the redshift $z_{*} = 1089.2$. 
			By considering this Planck measurement corresponding $\chi^2$ function can be defined.
			\item \textbf {BAO:}
			The baryon acoustic peak parameter $A$ best describes the distance-redshift relationship produced by a BAO measurement. The theoretical acoustic parameter can be defined in terms of the model parameters as \citep{N_2021},
			\begin{equation}
				A(H_0,\Omega_{m0}) = \frac{\sqrt{\Omega_{m0}}}{h(z_A)^{1/3}}\Big(\frac{1}{z_A}\int_{0}^{z_A}\frac{dz}{h(H_0,\Omega_{m0},z)}\Big)^{2/3}
			\end{equation}
			where $z_A$ is the redshift corresponding to the acoustic peak. To obtain the $\chi_{BAO}^2$, we compare the theoretical value with $A_{obs}=0.484\pm0.016$ at the redshift $z_A=0.35$ obtained from SDSS-BAO distance data \citep{10.1111/j.1365-2966.2011.19592.x}.
			\end{itemize}
			\begin{table*}
			\caption{The best estimated values and 68.3\% confidence limit for the $\Lambda$CDM model parameters.}
			\renewcommand{\arraystretch}{1.3}
			\begin{tabular}{lrrrrrrr}
				\hline\hline
				Datasets & $H_0$ & $\Omega_{m0}$ &   $M$ & $\chi^2$ & $\chi^2_{d.o.f}$ & AIC & BIC\\
				\hline\hline
				$H(z)$ & $71.592_{-0.991}^{+0.987}$ & $0.248_{-0.014}^{+0.015}$ &  ... & 29.746  &  0.5949 & -25.045 & -21.142 \\ 
				Pantheon & $69.991_{-3.401}^{+3.416}$ & $0.285_{-0.012}^{+0.013}$  &  $-19.357_{-0.108}^{+0.103}$ & 1035.680 & 0.9911 & -6.393 & 8.471 \\ 
				$H(z)$ + Pantheon + CMB & $69.276_{-0.572}^{+0.576}$ & $0.289_{-0.006}^{+0.006}$  &  $-19.377_{-0.017}^{+0.017}$ & 1074.862 & 0.9789 & -20.453 & -5.441 \\
				$H(z)$ + Pantheon + CMB + BAO & $69.257_{-0.567}^{+0.570}$ & $0.290_{-0.006}^{+0.006}$  &  $-19.378_{-0.017}^{+0.017}$ & 1074.929 & 0.9781 & -21.409 & -6.394 \\		
				\hline
			\end{tabular}
			\label{tab:params1}
		\end{table*}
		\begin{table*}
			\caption{The best estimated values and 68.3\% confidence limit for the PEDE model parameters.}
			\renewcommand{\arraystretch}{1.3}
			\begin{tabular}{lrrrrrrr}
				\hline\hline
				Datasets & $H_0$ & $\Omega_{m0}$ &   $M$ & $\chi^2$ & $\chi^2_{d.o.f}$ & AIC & BIC\\
				\hline\hline
				$H(z)$ & $73.578_{-1.024}^{+0.855}$ & $0.255_{-0.012}^{+0.014}$ &  ... & 33.697  &  0.6739 & -18.559 & -14.657\\ 
				Pantheon & $69.988_{-3.389}^{+3.402}$ & $0.332_{-0.012}^{+0.012}$  &  $-19.365_{-0.108}^{+0.103}$ & 1032.984 & 0.9885 & -9.125 & 5.739 \\ 
				$H(z)$ + Pantheon + CMB & $71.962_{-0.625}^{+0.625}$ & $0.281_{-0.006}^{+0.006}$  &  $-19.330_{-0.018}^{+0.018}$ & 1097.631 & 0.9997 & 2.626 & 17.638\\
				$H(z)$ + Pantheon + CMB + BAO & $71.966_{-0.618}^{+0.618}$ & $0.280_{-0.006}^{+0.006}$  &  $-19.330_{-0.018}^{+0.017}$ & 1097.642 & 0.9988 & 1.633 & 16.648\\		
				\hline
			\end{tabular}
			\label{tab:params2}
		\end{table*}
		\begin{figure}
			\centering
			\includegraphics[width=0.45\textwidth]{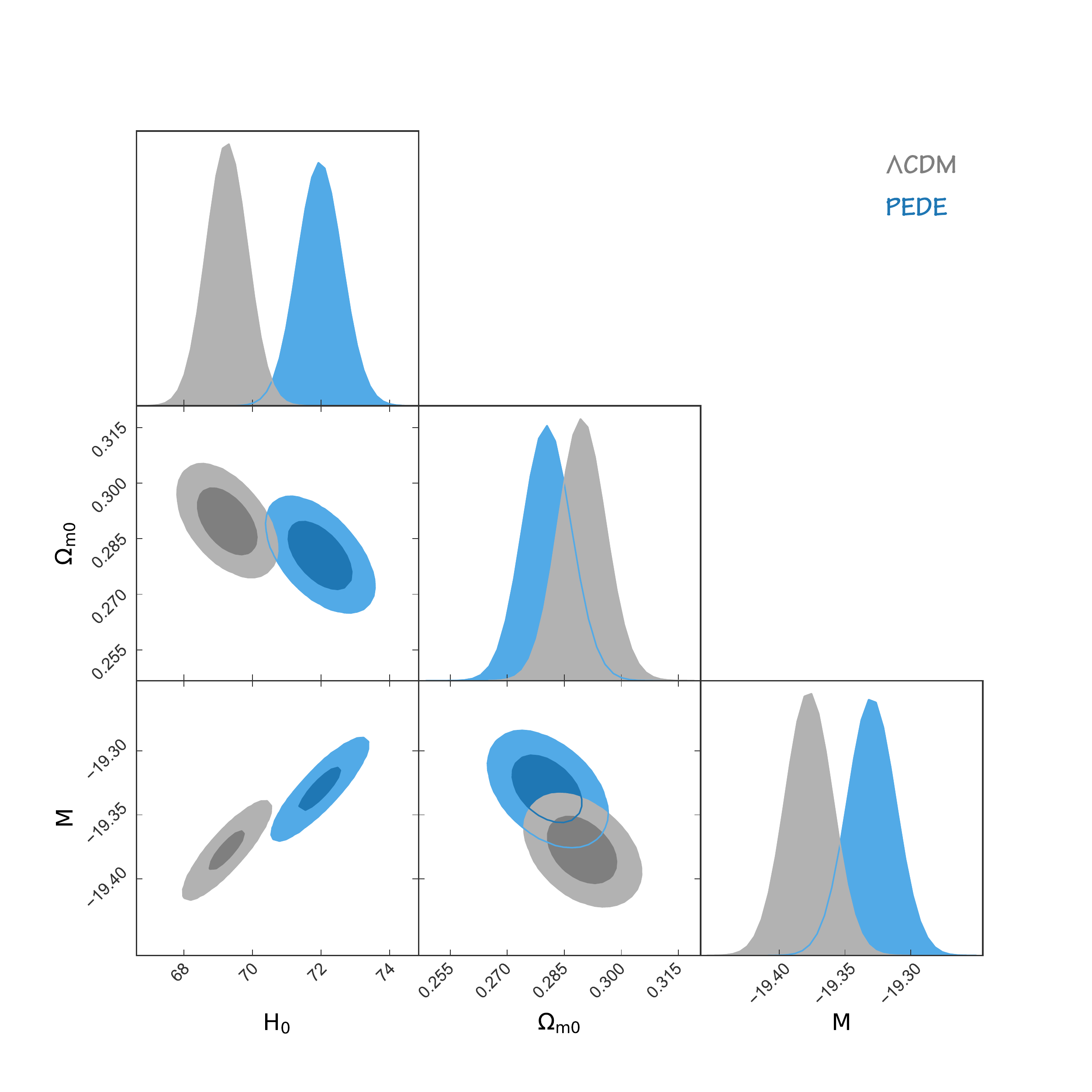}
			\caption{The two dimensional posterior contours with one dimensional posterior distribution (using pygtc open python package \citep{Bocquet2016}) from the $H(z)$+Pantheon+CMB+BAO datasets for the $\Lambda$CDM and PEDE model parameters}
			\label{fig:pygtc plot}
		\end{figure}
		\begin{figure}
			\centering
			\includegraphics[width=0.45\textwidth]{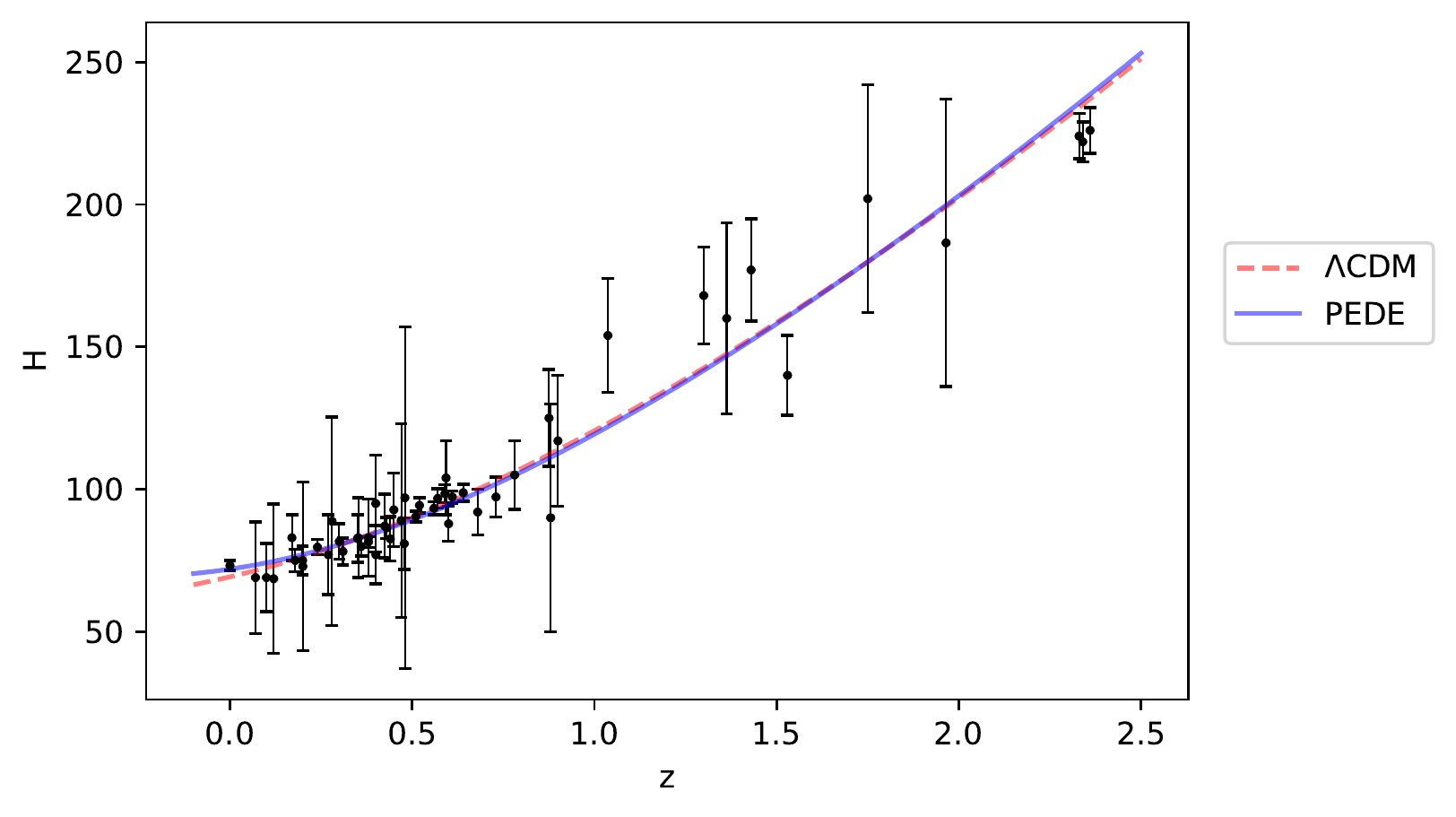}
			\caption{The evolution of Hubble parameter with redshift in $\Lambda$CDM and PEDE models compared with H(z) data points}
			\label{fig:Hfit}
		\end{figure}
		\begin{figure}
			\centering
			\includegraphics[width=0.45\textwidth]{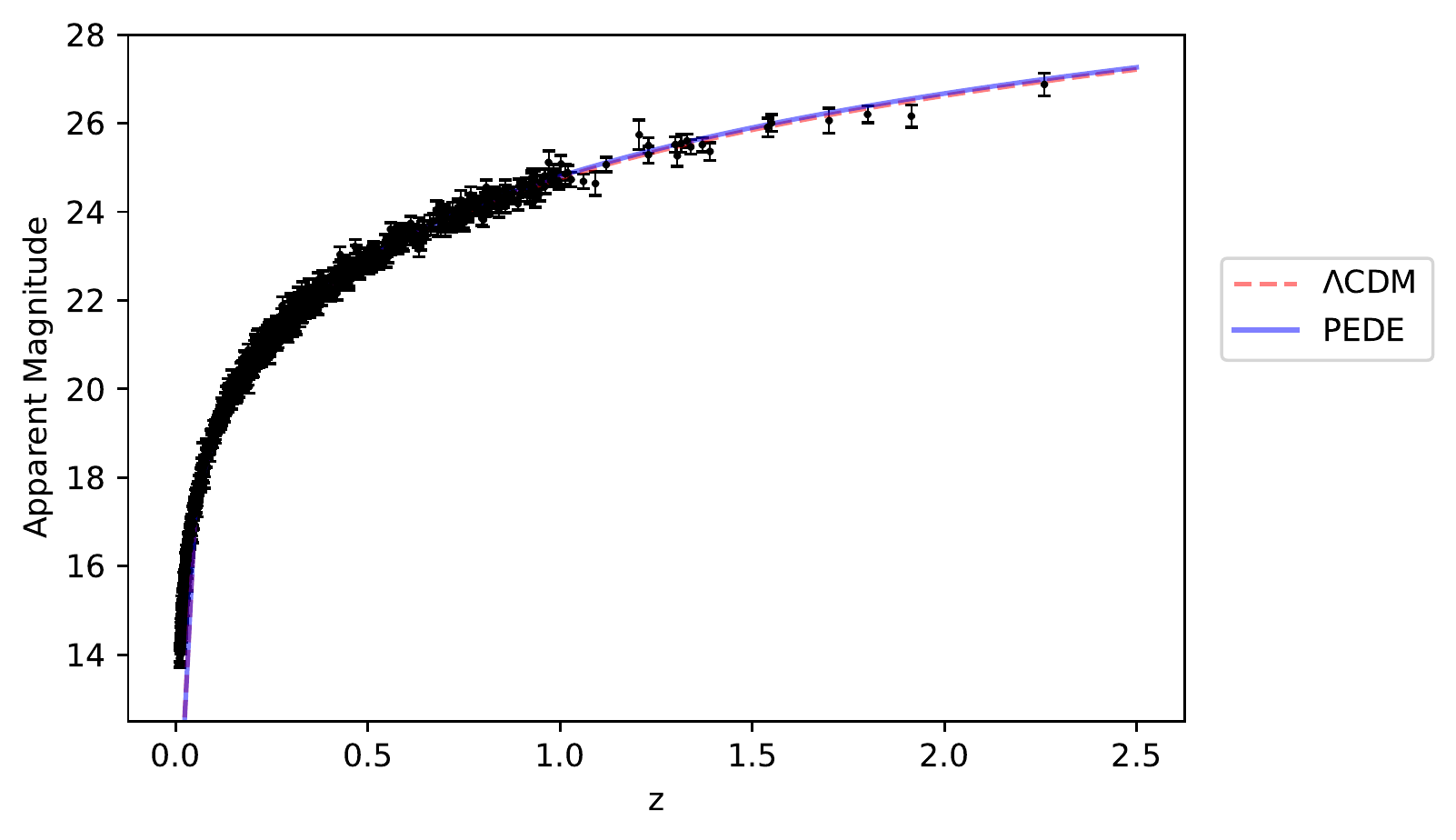}
			\caption{The evolution of apparent magnitude of supernovae with redshift in $\Lambda$CDM and PEDE models compared with Pantheon data points}
			\label{fig:Pfit}
		\end{figure}

		The best estimated parameters for the $\Lambda$CDM and PEDE model within 68.3\%  confidence level are given in Tab.\ref{tab:params1} and Tab.\ref{tab:params2} respectively and Fig.\ref{fig:pygtc plot} represents the two dimensional posterior contours and one dimensional posterior distribution for the best fit parameter values obtained for $H(z)$+SNIa+CMB+BAO dataset. The minimum $\chi^2$ value as well as $\chi^2_{d.o.f}$ are also mentioned in the tables. The $\chi^2$ for the combination of datasets is obtained by adding the individual $\chi^2$ minimum values for each dataset. The AIC and BIC values are also displayed in the table. In Fig.\ref{fig:Hfit} the evolution of the Hubble parameter for the PEDE and $\Lambda$CDM models using the best predicted model parameter values for the combined dataset are displayed and contrasted with the $H(z)$ data points. The evolution of apparent magnitude $m$ of supernovae for both the models are also compared with the Pantheon data points in Fig.\ref{fig:Pfit}.
		
		It is evident that the PEDE model can give a higher best fit value for the Hubble constant than $\Lambda$CDM except when the Pantheon sample is considered alone. For the Pantheon dataset the $\chi^2$ values for both models are similar. The addition of Supernovae data pushes the Hubble constant in PEDE model to relatively lower values. However, adding the $H(z)$ and CMB seems to increase the $H_0$ value and lower the $\Omega_{m0}$ value in the case of PEDE model whereas, for $\Lambda$CDM, the reverse is true. 	For the combined dataset of  $H(z)$+ Pantheon +CMB, the estimated Hubble constant is $69.257_{-0.567}^{+0.570}$ for $\Lambda$CDM and $71.966_{-0.618}^{+0.618}$ for PEDE with a $\chi^2$ difference of -22.713. The higher $H_0$ value from the PEDE model can be regarded as an indicator of its ability to resolve the tension in the Hubble constant.  Even though the dataset combination provides a higher $H_0$ value for the PEDE model, it happens at the cost of an increase in $\chi^2$, AIC and BIC values. 
		Similar problem of alleviation of tension by worsening the $\chi^2$ values compared to the $\Lambda$CDM model has been pointed out in Ref.\cite{pan2020reconciling}. The $\chi^2$ value listed in both tables points out the fact that $\Lambda$CDM is still favored over PEDE model when the datasets are considered together. 

		\section{Evolution of Cosmological Parameters}
		\subsection{Hubble parameter} 
\begin{figure}
			\centering
			\includegraphics[width=0.45\textwidth]{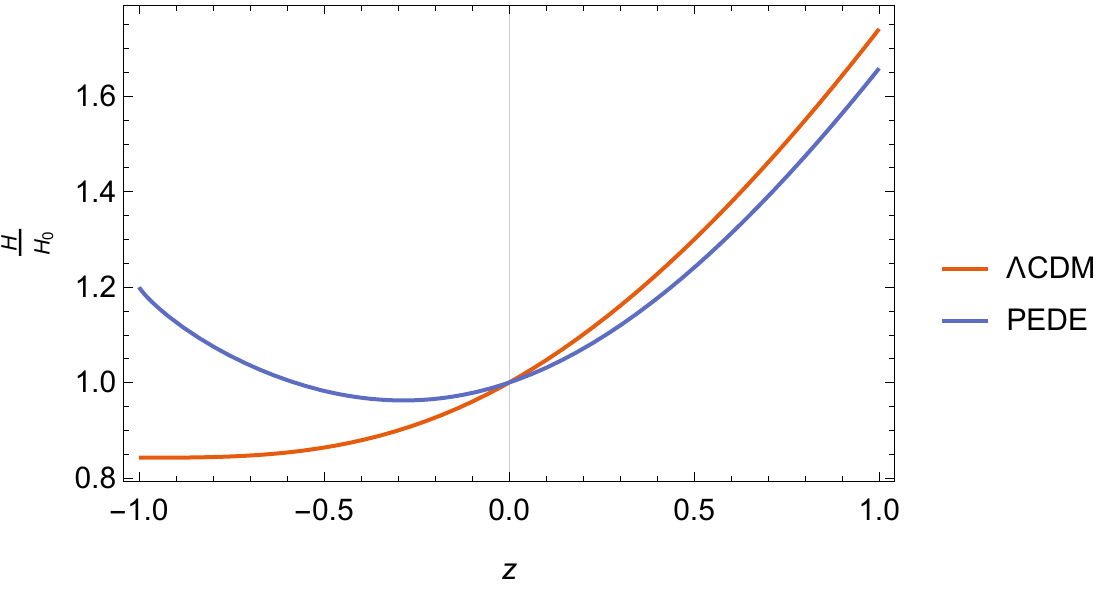}
			\caption{The evolution of $H/H_{0}$ with $z$}
			\label{fig:Hplot}
		\end{figure}

		The late phase evolution of the Hubble parameter in PEDE model for a flat universe can be expressed as,
		\begin{equation} \label{H1} 	
			H^2(z)= H_0^2\Big[\Omega_{m0}(1+z)^3+\Omega_{\scaleto{DE}{3.5pt},0}\big(1-\tanh\big[\log_{10}(1+z)\big]\big)\Big].
		\end{equation}
		where $\Omega_{\scaleto{DE}{3.5pt},0}=1-\Omega_{m0}$.
		The evolution of the normalized Hubble parameter,  $H(z)/H_0$ is plotted in Fig.\ref{fig:Hplot}, for both the standard as well as PEDE model with parameter values obtained for the dataset combination $H(z)$+Pantheon+CMB+BAO in the earlier section.
		It is evident from the figure that the evolution of the Hubble parameter in the present model has a substantial difference from the canonical model.  The Hubble parameter in $\Lambda$CDM is a smooth decreasing function of redshift that reaches a constant value in the future epoch. But in the present PEDE model, the Hubble parameter decreases until around $z\approx-0.3,$ which corresponds to a future time and then begins to increase. The dominating phantom component, which causes a faster acceleration in the universe's expansion, is the clear cause of the increase in the Hubble parameter. When $z\to -1$ the Hubble parameter in PEDE model asymptotically approaches $1.2 H_0$. 
		However, this non-monotonic behavior can eventually cause the 
		violation of the
		fundamental thermodynamic principle, as we will show in a later section. 
		\subsection{Deceleration Parameter}
\begin{figure}
			\centering
			\includegraphics[width=0.45\textwidth]{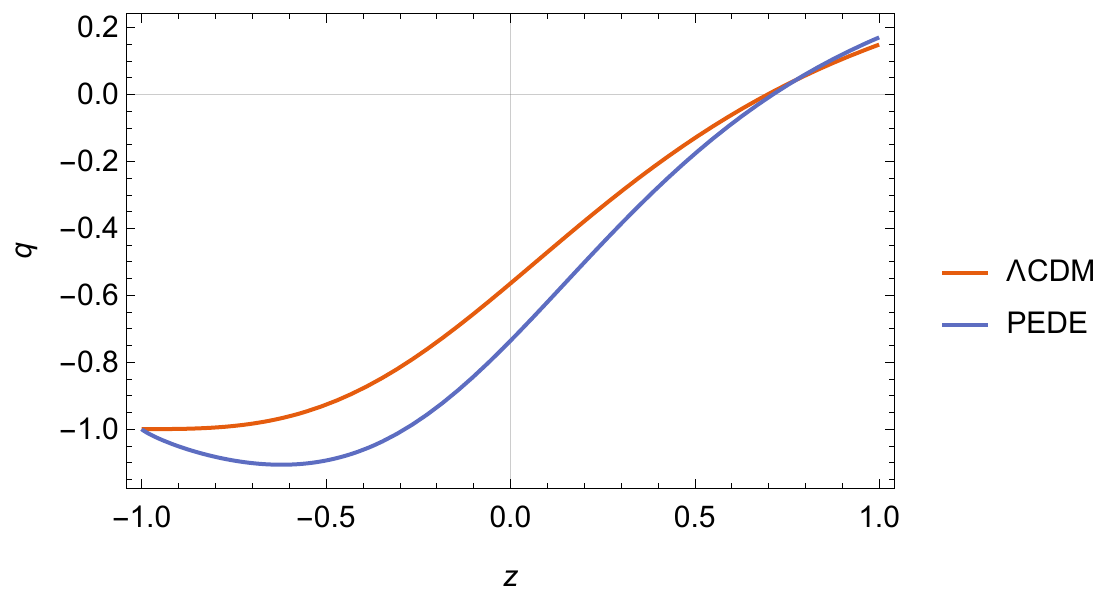}
			\caption{The evolution of $q$ with $z$}
			\label{fig:qplot}
		\end{figure}

		As the two quantities, the Hubble parameter and the deceleration parameter, are the focus of cosmology, we may now turn our attention to the evolution of deceleration parameter. For the PEDE model, we have observed a peculiar tendency in the evolution of the Hubble parameter, which also has an impact on the evolution of the deceleration parameter. The standard definition of the deceleration parameter is provided by,
		\begin{equation}  	
			q=-1-\frac{\Dot{H}}{H^2}.
		\end{equation}
		The over dot represents derivative over cosmic time. This equation can suitably be expressed in terms of a convenient variable,
		$x=\ln a,$
		\begin{equation}  	
			q=-1-\frac{1}{2h^2}\frac{dh^2}{dx}.
		\end{equation}
		where $h=H/H_0.$
		The evolution of deceleration parameter for both the canonical and PEDE models are
		plotted in Fig.\ref{fig:qplot}.
		As anticipated, the evolution of the deceleration parameter in PEDE model shows a marked deviation from the standard behavior.
		According to the standard model, the deceleration parameter decreases from its positive values and reaches -1 asymptotically. For $\Lambda$CDM, the transition redshift and the value of deceleration parameter for the present epoch (z=0) are found to be $z_*=0.70$ and $q_0=-0.56$ respectively. As shown in Fig.\ref{fig:qplot}, the deceleration parameter in PEDE model as well declines from the positive region and crosses into the negative region corresponding to the accelerating epoch at a redshift $z_*=0.71$. The current value of q is found to be, $q_0=-0.74$. However, below a redshift of $z\approx-0.3$, the $q$ parameter in PEDE model evolves to values less than -1 and attains a minimum, then increases and reaches the value -1 asymptotically. 
		This behavior is in line with that of both the Hubble parameter and dark energy density and is very unconventional as compared to the standard $\Lambda$CDM cosmology.
		
		\section{Energy conditions}
		Energy conditions are useful to analyze the physical feasibility as well as predicting the occurrence of singularities concerning a cosmological model. In principle, 
		energy conditions are expressing the fact that energy density measured in a region can't be negative. The general energy conditions in cosmology are obtained by limiting the energy momentum tensor, $T_{\mu\nu}$ on physical grounds \citep{hawking1973large,carroll2003can,santos2007energy,Bouhmadi_L_pez_2015}.
		\begin{itemize}
			\item \textbf{Strong energy condition (SEC)}\\
			The SEC states that $(T_{\mu\nu}-\frac{1}{2}Tg_{\mu\nu})u^\mu u^\nu \geq 0$ for all time like vectors $u^\mu$.
			SEC is satisfied by gravitationally interacting cosmic components. For a perfect fluid, it is equal as, 
			\begin{equation}
				\rho+3p\geq 0 
			\end{equation}
			For matter and radiation, the SEC is satisfied in every phase of evolution of universe. But dark energy, because of having a negative pressure doesn't hold this energy condition. The basic reason for the violation of this by dark energy can be attributed to its anti-gravitating property. This implies that SEC is satisfied in both radiation and matter dominated eras but violated in late accelerated epoch. SEC is violated in other scenarios as well, cosmological inflation is one among them.
			\item\textbf{Weak energy condition (WEC)}\\
			The energy momentum tensor obeys the inequality $T_{\mu\nu}u^\mu u^\nu \geq 0$ for any timelike $u^\mu$.
			It is a direct mathematical expression for: the matter-energy density measured by any observer must not be negative. 
			The condition can simply be expressed as,
			\begin{equation}
				\rho\geq 0
			\end{equation}
			This is the WEC, which is believed to be satisfied by all the cosmic components throughout the expansion of the universe.
			
			\item\textbf{Dominant energy condition (DEC)}\\
			The DEC can be defined as: 
			“for every timelike $u_\mu$, $T^{\mu\nu}u_\mu u_\nu \geq 0$ and $T^{\mu\nu}u_\mu$ is a non-spacelike vector”\citep{hawking1973large}.
			In addition to energy density being non-negative, DEC accommodates causality as well, i.e, the energy density can't flow with a speed faster than that of light and the condition can also be given as,
			\begin{equation}
				\rho\geq |p|
			\end{equation}
			DEC holds true for the standard $\Lambda$CDM model. But this condition is found to be violated in PEDE model during the dark energy dominated late epoch except at the asymptotic limit where $\omega_{DE}\to -1$ 
			\item\textbf{Null energy condition (NEC)}\\
			NEC is implied by SEC and WEC, and stated as $T_{\mu\nu}k^\mu k^\nu \geq 0$ for every null vector $k^\mu$ or,
			\begin{equation}
				\rho+p\geq 0
			\end{equation}
			We have analyzed the variation of  $\rho+p$ with redshift to check the NEC and the plot is given in Fig.\ref{fig:NEC}.
			\begin{figure}
				\centering
				\includegraphics[width=0.45\textwidth]{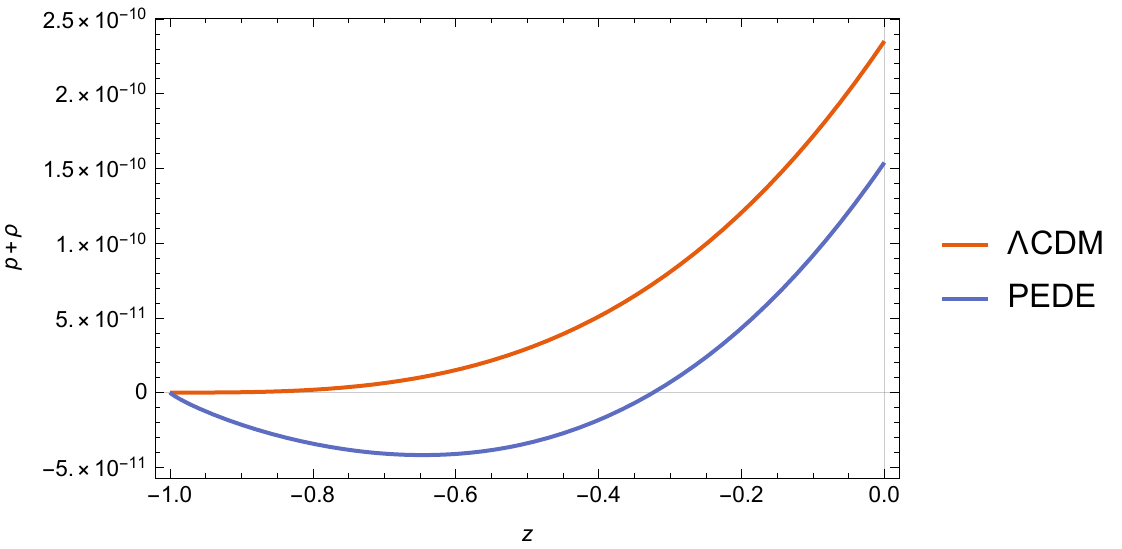}
				\caption{Violation of NEC shown by PEDE model}
				\label{fig:NEC}
			\end{figure}	
			It can be observed that NEC is satisfied throughout the evolution for $\Lambda$CDM model and in the case of PEDE model, it is satisfied during the initial part of evolution but is violated in the future, when $z \lesssim -0.3$. The plot indicates that, there occurs a chance for regaining the validity of NEC in the extreme future. However, we can say that $\rho+p$ is not greater than zero always, hence the model violates the NEC. 
			The violation of null energy condition is a general property of phantom cosmology which causes dark energy density to increase with expansion as suggested by the conservation equation. The violation of  NEC exhibited by phantom cosmologies usually leads to future singularities. However, in PEDE model, the dark energy density approaches a constant value rather than increasing boundlessly, there won't be any sudden singularities, but still the system may get unstable against perturbations.
		\end{itemize}
		
		\section{Statefinder analysis}
		The presence of a large number of dark energy models necessitates the comparison of them with the standard model and also among themselves. Such a comparison is needed to measure the advantage of the model over the standard model or any other existing models.  In 2003, Sahni et al.\cite{sahni2003statefinder} introduced a diagnostic tool to facilitate the comparison of cosmological models, using what is known as statefinder variables, represented as $\{r,s\}$. These are dimensionless geometric variables, constructed from the higher order derivatives of scale factor with cosmic time \citep{alam2003exploring}. 
		
		Among statefinder diagnostic pair, $r,$ is called the jerk parameter, which is defined as,
		\begin{equation}
			\label{sf-r}
			r=\frac{\dddot a}{aH^3}
		\end{equation}
		In comparison to the Hubble and deceleration parameters, this parameter becomes crucial in highlighting the difference between a model and any other model since it contains the triple derivative of the cosmic scale factor.
		The second parameter in the statefinder set $s,$ is constructed by combining the parameters $q$ and $r$, defined as,
		\begin{equation}
			\label{sf-s}
			s=\frac{r-1}{3(q-{1/2})} 
		\end{equation}
		We can rewrite the equations for $r$ and $s$ in terms of $h.$
		As we did earlier in the case of deceleration parameter, with a change of variable from $t$ to $x$, the Eqs. \ref{sf-r} and \ref{sf-s} can be rewritten as,
		\begin{equation}
			r  =\frac{1}{2h^2}\frac{d^2h^2}{dx^2}+\frac{3}{2h^2}\frac{dh^2}{dx}+1    
		\end{equation}
		\begin{equation}
			s= -\left(\dfrac{\dfrac{1}{2h^2}\dfrac{d^2h^2}{dx^2}+\dfrac{3}{2h^2}\dfrac{dh^2}{dx}}{\dfrac{3}{2h^2}\dfrac{dh^2}{dx}+\dfrac{9}{2}}\right).
		\end{equation}
		\begin{figure}
			\centering
			\includegraphics[width=0.45\textwidth]{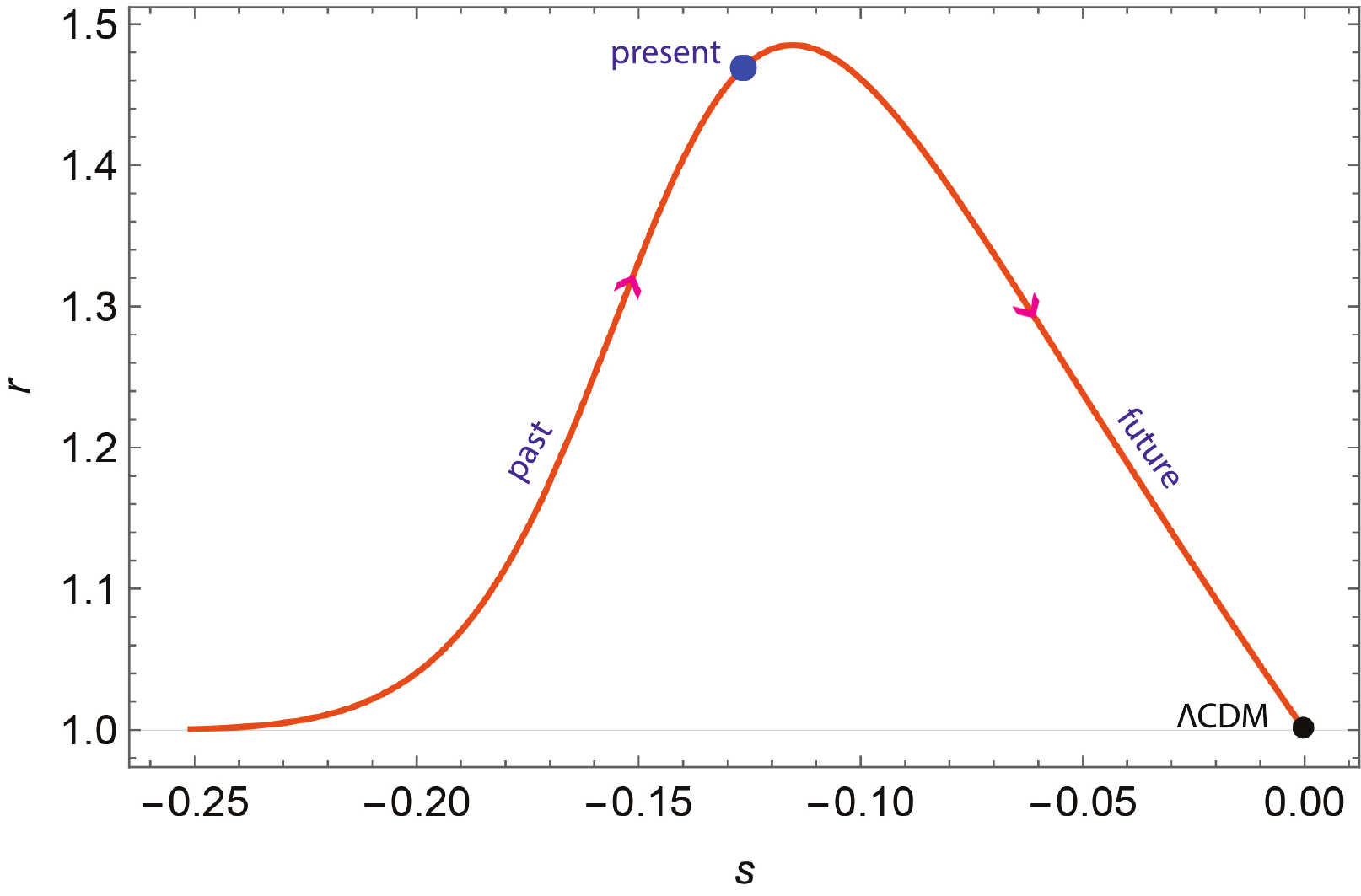}
			\caption{The evolution of statefinder pair}
			\label{fig:rsplot}
		\end{figure}
		The evolution of the present model in the statefinder plane is shown in Fig.\ref{fig:rsplot}. For reference, note that, 
		for $\Lambda$CDM model, the statefinder pair has a constant value of $\{r,s\}=\{1,0\}$ and therefore any departure from this point signifies the deviation from $\Lambda$CDM model. The $r$ parameter increases initially in the PEDE model before starting to decline and becoming closer to 1. On the other hand, the $s$ parameter monotonically rises to 0.
		The present values of statefinder parameters for the PEDE model are, $\{r_0,s_0\}=\{1.469, -0.126\}$, making the model clearly distinguishable from the standard model. It can be inferred from Fig.\ref{fig:rsplot}, the model approaches the $\Lambda$CDM values of statefinder in the far future, i.e., when the scale factor goes to infinity. However, during the entire evolution of the model, until $z\to-1,$  evolution of the parameters satisfies, $r>1$ and $s<0,$ which implies a Chaplygin gas like behavior with phantom nature. There are examples of dark energy models in the literature which show Chaplygin gas behavior with phantom nature \citep{wu2007modified, sadjadi2010crossing}. So the predominant behavior of the model is phantom-like throughout its evolution.

		\section{Thermodynamics of PEDE model 
		}
		
		In this section, we will analyze the thermal evolution of the model. We first consider the status of the generalized second law of thermodynamics and then the status of the principle of maximization of the entropy. 
		\subsection{Horizon entropy}
\begin{figure}
			\centering
			\includegraphics[width=0.45\textwidth]{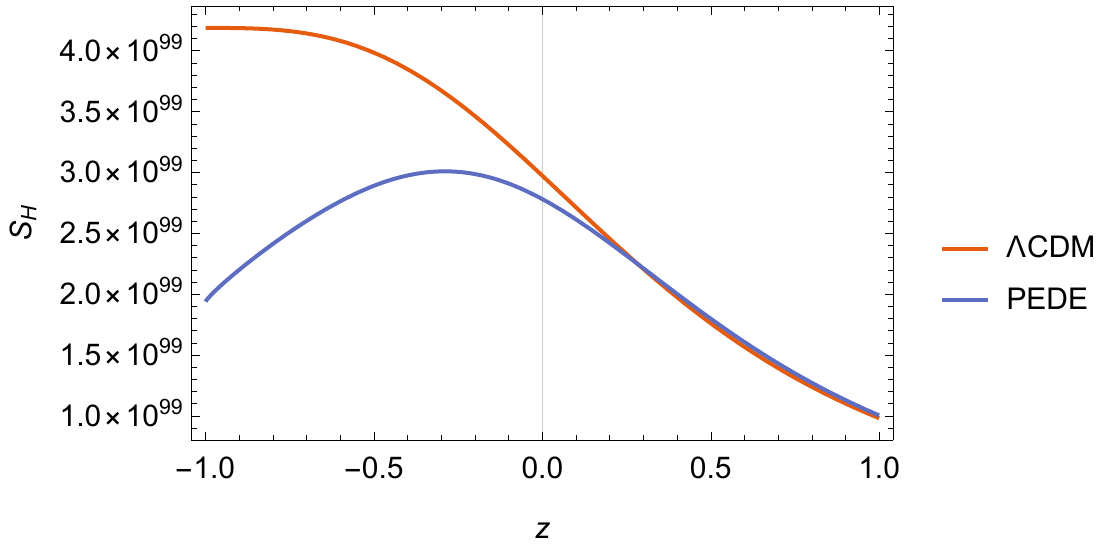}
			\caption{The evolution of Horizon entropy with redhsift}
			\label{fig:entroevol}
		\end{figure}

\begin{figure}
			\centering
			\includegraphics[width=0.45\textwidth]{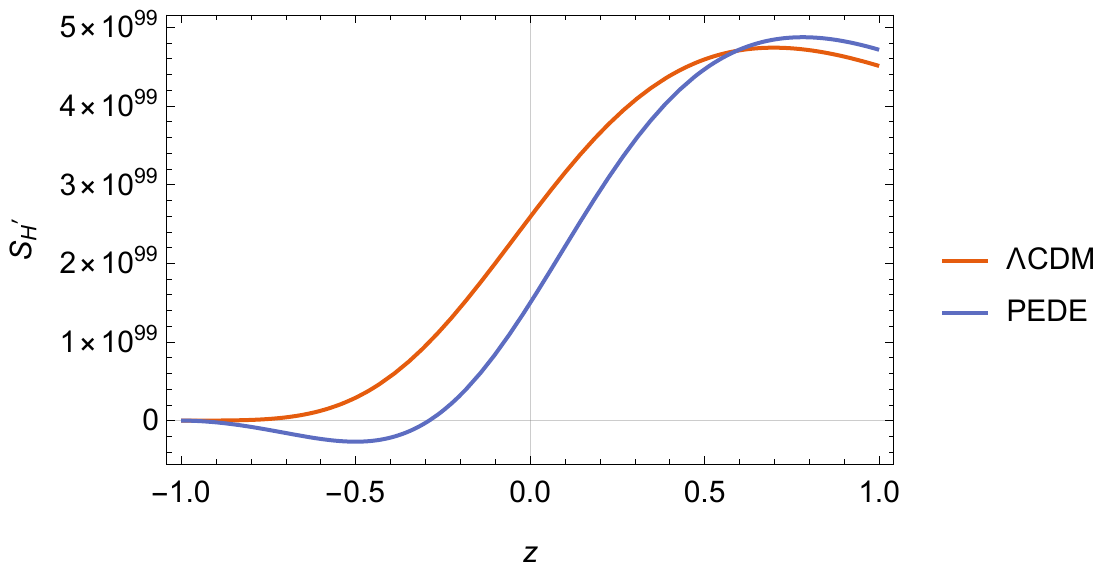}
			\caption{The evolution of the first derivative of Horizon entropy}
			\label{fig:entroder}
		\end{figure}

		The second law of thermodynamics constrains the evolution of entropy, as it states that: “the total entropy of the system should always increase".
		The law also holds true on cosmological scales.
		The Hubble horizon can be thought of as the thermodynamic boundary of the universe \citep{Davies_1987, Davis_2003}.  By taking account of horizon entropy,  the generalized second law of thermodynamics can be stated as \citep{setare2006generalized, mohan2020dynamical},
		\begin{equation}
			S_H'+S_m'\geq 0
		\end{equation}
		where $S_H$ is the entropy of the Hubble horizon and $S_m$ is the entropy of matter contained within the horizon. The prime denotes derivative with respect to an appropriate variable like cosmic time or scale factor. Since the matter entropy is several orders of magnitude less than the horizon entropy\citep{Egan_2010}, total entropy can be approximated to that of the horizon. The cosmic horizon entropy has a form similar to black hole event horizon entropy suggested by Bekenstein \cite{PhysRevD.7.2333} and is defined as,
		\begin{equation}
			\label{ent hori}
			S_H=\frac{A_Hk_B}{4l_P^2}
		\end{equation}
		where $A_H=4\pi r_H^2,$ the area of the horizon surface , $k_B$ is Boltzmann constant and $l_p$ is the Planck length. $r_H$ is the apparent horizon radius and for a spatially flat universe, $r_H=\frac{c}{H}$.
		Expressed in terms of $h$, the Eq.\ref{ent hori} becomes,
		\begin{equation}
			S_H=\frac{\pi c^5k_B}{\hslash G H_0^2 h^2}
			\label{eqn:hentro}
		\end{equation}
		Substituting the Hubble parameter of the PEDE model in the above equation, the evolution of the horizon entropy can be studied. Fig.\ref{fig:entroevol} presents the plot of its evolution with redshift. For better comparison, we have shown the entropy evolution of $\Lambda$CDM model in the same figure.
		The figure indicates that, in the PEDE model, the entropy increases first, attains a maximum value at $z\approx-0.3,$ and then decreases thereafter.  This, violates the generalized $2^{nd}$ law, according to which entropy will never decrease. 
		To add more clarity, let's consider the derivative of horizon entropy with respect to the scale factor. For an increasing entropy evolution, the derivative must be non-negative.
		\begin{equation}\label{entrorate}
			\begin{split}
				S_H'=&\frac{dS_H}{da}\\
				=&-\frac{\pi c^5 k_B }{\hslash GH_0^2}\;\frac{e^{-x}}{h^4}\left(\frac{dh^2}{dx}\right)
			\end{split}
		\end{equation}
		This shows that since $dh^2/dx<0$ up to $z \approx -0.3$, the rate of entropy, $S_H^{\prime}>0,$ which implies an initial increase in entropy. But in future evolution, where the redshift becomes less than around $-0.3$ the rate of change of Hubble parameter is such that, $dh^2/dx > 0,$ corresponding to which the entropy will decrease.
		The evolution of $S_H'$ is plotted against redshift (Fig.\ref{fig:entroder}).
		\noindent 
		The generalized second law of thermodynamics is violated as indicated by the drop in horizon entropy and its negative derivative in the later stages of evolution. 
		It is to be noted that, the redshift at which the entropy starts decreasing is the same as that of NEC violation.
		\subsection{Maximization of entropy}
\begin{figure}
			\centering
			\includegraphics[width=0.45\textwidth]{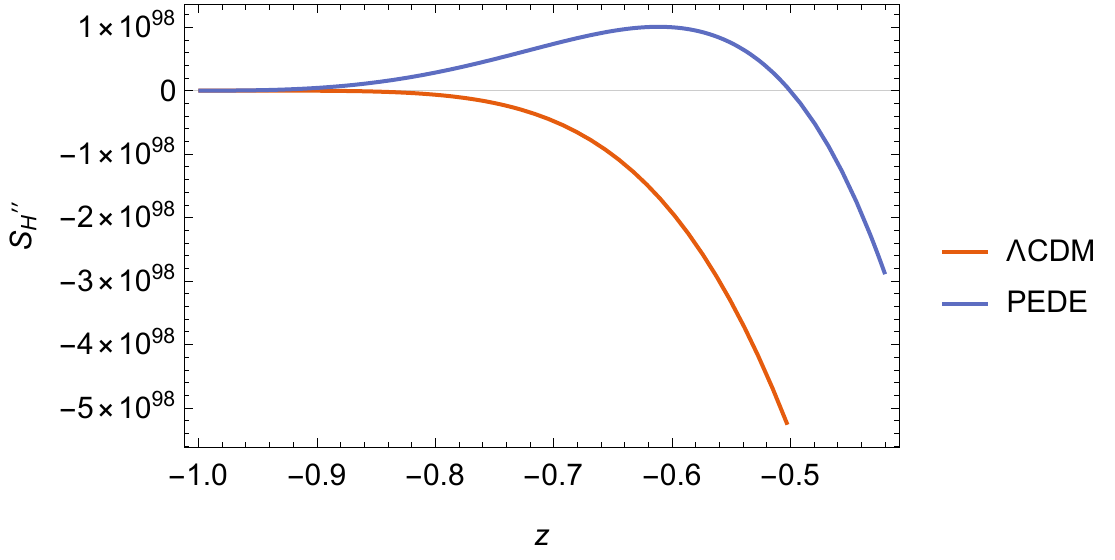}
			\caption{The evolution of the second derivative of Horizon entropy}
			\label{fig:ddentro}
		\end{figure}

		We plan to examine the entropy maximization in order to provide further insight into the thermal stability of the PEDE model.
		As already known any isolated system evolves to a state of equilibrium, at which the entropy is maximum. The condition, the convexity condition, to be satisfied by a system, to achieve a final maximum entropy state is that \citep{article}, 
		\begin{equation}
			S'' < 0 \hspace{0.5 cm} ;  \hspace{0.5 cm} \text{at least in the long run}.
		\end{equation}
		That is the second derivative of the total entropy with respect to any convenient cosmological variable, must be less than zero, at least in the end stage of the evolution. This otherwise suggests that the entropy must be a convex function of the said variable. The standard $\Lambda$CDM model is found to satisfy this condition at its end de Sitter epoch \citep{krishna2017holographic}. 
		To check the condition for the present model, second derivative of total entropy is required. With the earlier assumption of total entropy can be approximated to horizon entropy, the condition becomes $S_H'' < 0,$ where the derivative can be obtained with respect to the scale factor of the evolution of the universe. Differentiating Eq.\ref{entrorate} gives,
		\begin{equation}
			\begin{split}
				S_H'' &= \frac{dS_H'}{da} \\
				&= -\frac{\pi c^5k_B}{\hslash GH_0^2}\;\frac{e^{-2x}}{h^4}\left[\frac{d^2h^2}{dx^2}-\frac{dh^2}{dx}-\frac{2}{h^2}\left(\frac{dh^2}{dx}\right)^2\right]
			\end{split}		
		\end{equation} 
		We found the evolution of  $S_H''$ which is shown in Fig.\ref{fig:ddentro}.
		\noindent  It is evident from the figure that the second derivative of the entropy in PEDE scenario increases first, attains a maximum, and then decreases approaching zero from above. This is a clear violation of the convexity condition.
		This implies that the entropy evolution of PEDE model doesn't behave like that of an ordinary thermodynamic system\citep{article}. This boundlessness in entropy can cause dynamical instabilities in the future evolution of the universe. 
		
		To sum up, the thermal evolution of the PEDE model shows two major issues. First, it violates the generalized second law of thermodynamics and second, the entropy is not bounded.  These may doom the plausibility of this dark energy model in comparison with the standard $\Lambda$CDM model.

		\section{Dynamical system analysis}
		
		In the previous section, we analyzed the thermal evolution of the PEDE model of dark energy and it reveals that the model is breaking the conventional thermodynamics. One can analyze the dynamical system behavior of the model to get what it actually  implies at the end stage of the universe. Even though we may get a glimpse of this information from the Hubble parameter evolution, we are performing here the dynamical system analysis to analyze exactly the dynamical behavior of the model in the asymptotic limits
		\citep{wainwright1989dynamical,mathew2022running}. 
		
		To facilitate the analysis, we define the simple dimensionless dynamical variables,
		\begin{equation}
			u=\frac{\rho_m}{3H^2} \hspace{0.2cm},\hspace{0.2cm} v=\frac{\rho_{\scaleto{DE}{3.5pt}}}{3H^2}
		\end{equation}
		We require a few basic equations to analyze the dynamical system's behavior using the differential equations on $u$ and $v$.
		The first Friedmann equation (Eq.\ref{F1}),taking $8\pi G= c=1$, can be expressed as, 
		\begin{equation}
			3H^2=\rho_m+\rho_{\scaleto{DE}{3.5pt}}
		\end{equation}
		which makes $u+v=1.$
		From the conservation equations for non-interacting matter and dark energy (Eq.\ref{cont}),
		\begin{equation}
			\begin{split}
				\frac{d\rho_m}{dx}&=-3\rho_m \\
				\frac{d\rho_{\scaleto{DE}{3.5pt}}}{dx}&=-3\rho_{\scaleto{DE}{3.5pt}}(1+\omega_{\scaleto{DE}{3.5pt}}) 
			\end{split}
		\end{equation}
		The autonomous coupled differential equations can be obtained from the above equations as,
		\begin{equation}
			\begin{split}
				\label{d1}
				\frac{du}{dx} &= -3u+3u\Big[u+v(1+\omega_{\scaleto{DE}{3.5pt}})\Big]=f(u,v) \\
				\frac{dv}{dx} &= -3v(1+\omega_{\scaleto{DE}{3.5pt}})+3v\Big[u+v(1+\omega_{\scaleto{DE}{3.5pt}})\Big] =g(u,v)
			\end{split}
		\end{equation}
		
		The properties of the critical points, which indicate the equilibrium points, are of relevance here. To find the critical points, we set $f(u,v)=0$ and $g(u,v)=0$. The resulting points are $(u_c,v_c)=(1,0),(0,0),(0,1)$. It can be shown that the point (1,0) corresponds to matter dominated epoch while (0,1) corresponds to the dark energy dominated asymptotic future epoch. The (0,0) point is a trivial solution and that implies a universe with no matter and dark energy which is called an empty universe or Milne universe. The point (0,0) is not taken into more consideration as it yields a solution that does not have significance in the present context.
		
		To study the stability of the other critical points, the method of linear perturbation is applied i.e., $u\to u'=u_c+\delta u$ and $v\to v'=v_c+\delta v,$ where $\delta u$ and $\delta v$ are infinitesimally small deviations from the equilibrium values. Linearizing the system of differential Eq.\ref{d1}, with respect to the perturbation, we can obtain a matrix equation of the form,
		\renewcommand\arraystretch{2.5}
		\begin{equation}
			\begin{bmatrix}
				\delta u' \\
				\delta v' 
			\end{bmatrix}
			=\begin{bmatrix}
				\left(\dfrac{\partial f}{\partial u}\right)_c & \left(\dfrac{\partial f}{\partial v}\right)_c\\
				\left(\dfrac{\partial g}{\partial u}\right)_c & \left(\dfrac{\partial g}{\partial v}\right)_c
			\end{bmatrix}
			\begin{bmatrix}
				\delta u \\
				\delta v
			\end{bmatrix}
		\end{equation}
		The $2\times 2$ matrix on the right hand side is called the Jacobian matrix. The Jacobian matrix obtained for the PEDE model is given as,
		\begin{equation}
			\begin{bmatrix}
				-3+6u+3v(1+\omega_{\scaleto{DE}{3.5pt}}) & 3u(1+\omega_{\scaleto{DE}{3.5pt}})\\
				3v & 3u+(1+\omega_{\scaleto{DE}{3.5pt}})(6v-3)
			\end{bmatrix}
		\end{equation}
		The eigenvalues can be found by diagonalizing the matrix and then the stability of the system depends upon the sign of eigenvalues. If both the eigenvalues evaluated for the critical point are negative, the system is stable. The phase space trajectories, irrespective of their origin will converge to that critical point. The system is said to be unstable if both the eigenvalues are positive, and hence the trajectories will diverge away from the critical point. If one of the eigenvalues is positive and the other is negative, the system is said to be in saddle at that critical point. 
		\begin{figure}
			\centering
			\includegraphics[width=0.45\textwidth]{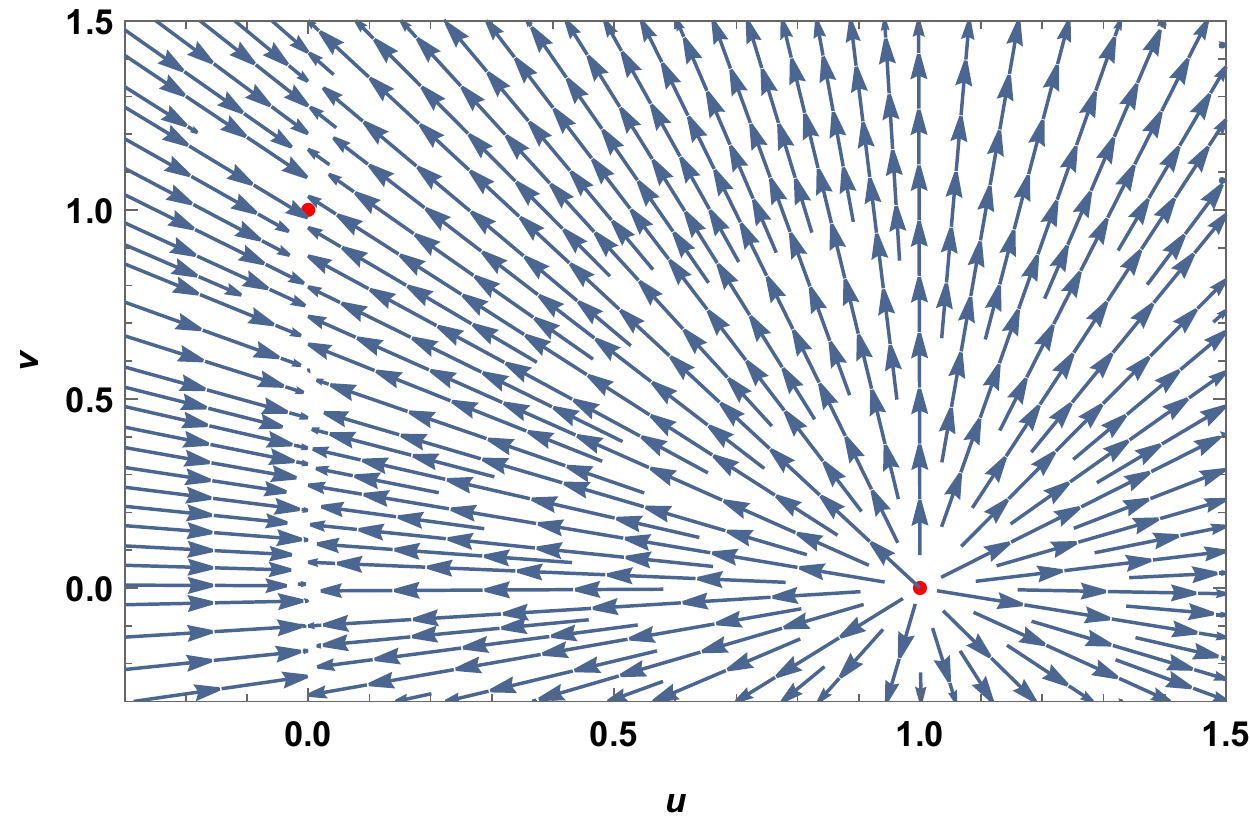}
			\caption{The phase space diagram corresponding to $\omega_{\scaleto{DE}{3.5pt}}(z=-1)=-1$}
			\label{fig:phase future}
		\end{figure}
		For PEDE model, the eigenvalues corresponding to the critical point (1,0) are obtained as ($3$ ,\;$-3\omega_{\scaleto{DE}{3.5pt}}$). Since the equation of state of dark energy is found to be less than -1, it turns out that, both the eigenvalues will become positive, making the system unstable in the matter dominated epoch. The instability drives to further expansion of the universe. 
		The second critical point (0,1), refers to the future state and has eigenvalues,  $(3\omega_{\scaleto{DE}{3.5pt}}, 3(1+\omega_{\scaleto{DE}{3.5pt}}))$. In the asymptotic state, which corresponds to the de Sitter epoch, the equation of state of dark energy becomes  $-1.$ It will lead to a set of eigenvalues $(-3,0$),  with one being negative while the other is zero. The nature of this critical point is not trivial.
		Let us now look at the phase space trajectory, given in Fig.\ref{fig:phase future} for these critical points.  It suggests that the critical point (1,0) indeed is a source point,  as the trajectories are diverging from it. While regarding the second equilibrium point, (0,1),  it is seen that the trajectories appear to converge to a line, rather than to a point, parallel to the $v$ axis corresponding to the dark energy in the dynamical system plane.  To be specific, the primary feature of the `stable point' is that, the surrounding trajectories must converge to that point. Since that feature is not strictly obeyed here, there is no guarantee that the critical point is a stable one.  At the same time, in an infinitesimal region around the point (0,1) the trajectories can be considered converging. Let us analyze a bit further regarding the nature of the critical point (0,1). There are papers in the literature that argues, if one of the eigenvalues of a critical point is zero, the nature of that can be judged from the sign of the other eigenvalue \citep{banerjee2015stability}. However, it is to be noted from the 
		thermodynamic analysis, that 
		the entropy is not maximizing in the asymptotic future corresponding to the critical point (0,1). The decreasing entropy as observed in Fig.\ref{fig:entroevol} makes this state of the system thermodynamically unstable in the future epoch and correspondingly the critical point (0,1) can said to be thermally unstable. But  
		the dynamical system analysis based on the phase space diagram is suggesting that the system is at least locally stable. That is we come across a situation where the critical point (0,1) is thermodynamically unstable but dynamically it seems to be apparently stable at least in the local sense. To clarify this apparent paradoxical situation further we will analyze the behavior of sound speed in the following subsection.

		\subsection{Dark energy sound speed} 
\begin{figure}
			\centering
			\includegraphics[width=0.45\textwidth]{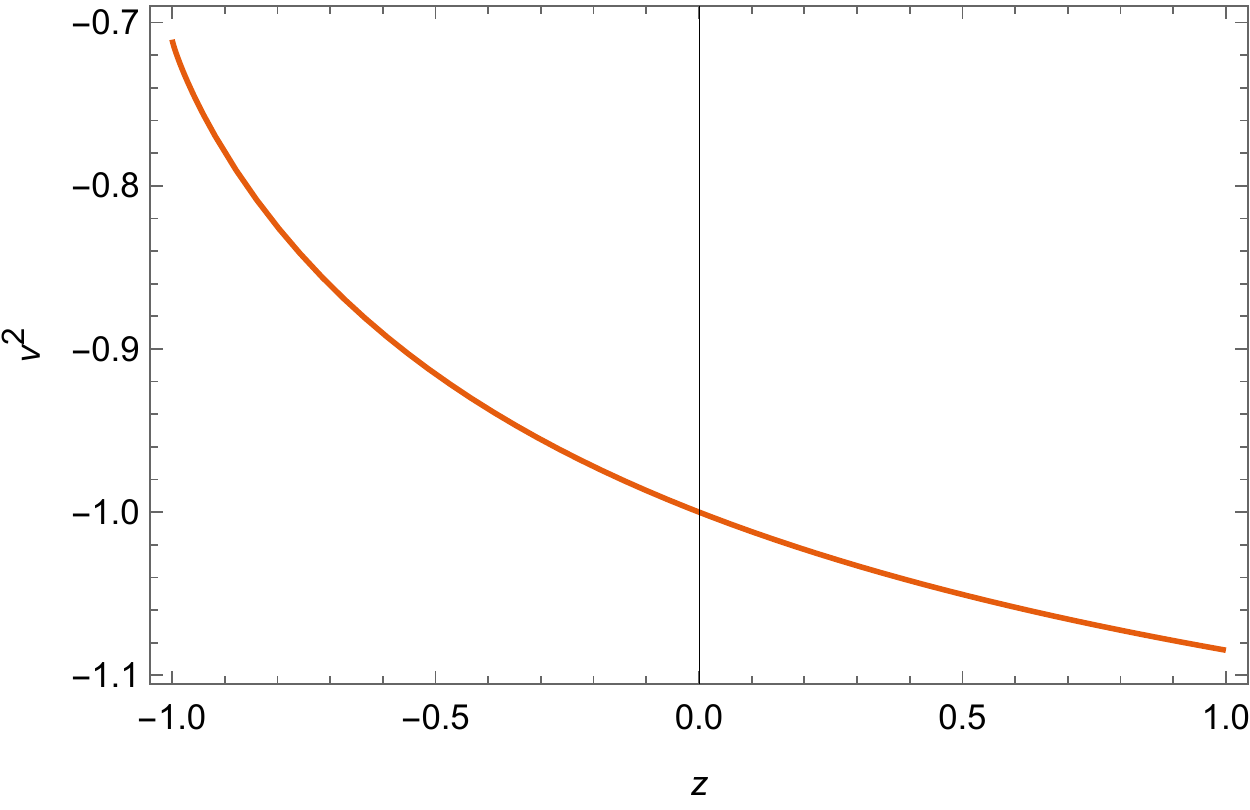}
			\caption{The evolution of squared sound speed with redshift}
			\label{fig:stability}
		\end{figure}

		The dark energy sound speed is described by the ratio of pressure to density variations as\citep{https://doi.org/10.48550/arxiv.2001.02368},
		\begin{equation}
			v_s^2=\dfrac{dp_{\scaleto{DE}{3.5pt}}}{d\rho_{\scaleto{DE}{3.5pt}}}
		\end{equation}
		It has been argued that the physical values of  $v_s^2$ ought to be in the range from 0 to 1. Outside of this region, one encounters gradient/tachyonic instabilities and/or instabilities linked to superluminal propagation\citep{Vagnozzi_2020}. A real value of sound speed shows a regular propagating mode for density perturbation while an imaginary value corresponds to an irregular wave equation implying a classical instability of a given perturbation. Since a cosmological constant suffers no spatial perturbations and cosmology with dark energy in the form of a cosmological constant thus has no sensitivity to the dark energy sound speed, it is vital to keep in mind that the study is only applicable when the equation of state of dark energy, $\omega_{\scaleto{DE}{4pt}}\neq -1$\citep{https://doi.org/10.48550/arxiv.2001.02368}. One can rewrite the equation for $v_s^2$ as,
		\begin{equation}
			v_s^2=\dfrac{\dot{p}_{\scaleto{DE}{3.5pt}}}{\dot{\rho}_{\scaleto{DE}{3.5pt}}}
		\end{equation}
		where $\dot{p}_{\scaleto{DE}{3.5pt}}=\omega_{\scaleto{DE}{3.5pt}}\dot{\rho}_{\scaleto{DE}{3.5pt}}+\dot{\omega}_{\scaleto{DE}{3.5pt}}\rho_{\scaleto{DE}{3.5pt}}$.  
		Substituting with PE dark energy, the above equation reduces to,
		\begin{equation}
			v_s^2=-\Big(\dfrac{2\tanh\big[\log_{10}(1+z)\big]}{3\ln10}+1\Big)
		\end{equation}
		When $z \rightarrow-1$, $v_s^2$ reaches to a value $\approx -0.7$. The Fig.\ref{fig:stability} gives the complete evolution of $v_s^2$.
		It is evident from the figure that the value of  $v_s^2$ is negative including the far future epoch marking instabilities at the perturbation level. The perturbation has become an irregular wave equation. This negative squared speed shows an exponentially growing mode for a density perturbation. In other words, an increasing density perturbation induces a lowering pressure, supporting the emergence of instability\citep{https://doi.org/10.48550/arxiv.2001.02368,KIM2008118,Myung_2007}.
		Moreover, the increment of instability is inversely proportional to the wavelength of the perturbations\citep{armendariz1999k}, and therefore the PEDE model for which $v_s^2<0$ is asymptotically unstable. Previously when we analyzed the stability conditions corresponding to the critical point (0,1), we found that the corresponding state seems to be stable only locally. The thermal instability of the same state compelled us to investigate further. In this regard, the acoustic speed evolution now shows that this asymptotic state is not stable. Hence taking account of the thermal evolution and evolution of the sound speed we have to conclude that the critical point (0,1) may be unstable.

		Before concluding the section we would like to mention a recent work in which the authors have performed the dynamical stability analysis of the very same model. Hern\'{a}ndez-Almada et al.\cite{hernandez2020generalized} have studied the dynamical stability of the PEDE model of dark energy, in a three-dimensional plane with variables, $T=H_0/(H_0+H), \,  \Omega_m = H_0^2\Omega_{m0}/(a^3 H^2), \, \Omega_\gamma=H_0^2\Omega_{\gamma0}/(a^4 H^2)$. Interestingly they obtained the eigenvalues corresponding to the future epoch as all negative, implying a stable nature and from the 3D phase diagram (Fig.7 in \citep{hernandez2020generalized}), the trajectories appear to be converging to a point in T axis. It is also to be noted that, the authors\citep{hernandez2020generalized}, predict an effect of the PE dark energy in the very early stage of the evolution of the universe, such that it can even drive the early acceleration equivalent to inflation. This argument is in fact contrary to the claim of originators of this very model, that the PEDE dark energy will emerge as the universe expand such that it can't have any commendable effect on the early stage of evolution. Moreover, they haven't studied the thermal evolution of the model. However, it is essentially difficult to comprehend the dynamical stability of a system that is thermally unstable. 
		

		\section{Conclusions }
		
		In this work, we explore the background evolution of the cosmological model, with the phenomenologically emergent dark energy. We analyzed the evolution of basic parameters and the status of the model using statefinder diagnostic. We also examined the thermodynamic evolution and dynamical system behavior of the model. 
		
		Major observations derived from our study are given below.
		\begin{itemize}
			\item Unlike the standard model, the PEDE model offers variable dark energy that grows as the universe expands. The evolution of the equation of state shows that the dark energy exhibits phantom behavior ($\omega_{\scaleto{DE}{4pt}}<-1$) till the asymptotic future epoch. Due to the phantom energy, the Hubble parameter gets shifted to higher values for the present epoch thereby alleviating the Hubble tension.
			The parameter values obtained for the PEDE model from our analysis are: $H_0=71.966\pm0.618\ \si{ kms^{-1}Mpc^{-1}}$ and $\Omega_{m0}=0.280\pm0.006$ for the data combination $H(z)$+Pantheon+CMB+BAO.
			
			\item The evolution of Hubble and deceleration parameters are found to be much different from the standard model.
			Hubble parameter decreases initially, as seen in Fig.\ref{fig:Hplot}, but after $z\approx-0.3$, it appears to be increasing. The increase in $H$ during the late phase is due to the dominating phantom fluid which is not expected in the standard cosmological scenario.
			
			The evolution of the deceleration parameter also reflects the divergent behavior of the Hubble parameter and equation of state (see Fig.\ref{fig:qplot}). Like the standard model, the PEDE model predicts an earlier decelerated epoch dominated by matter and a later accelerated period dominated by dark energy. The redshift of the transition from decelerated to accelerated epoch is determined as $z\approx0.7$. However, owing to the growth of phantom dark energy density in PEDE model, the value of $q$ in the future decreases from -1 at $z\approx-0.3$ and eventually returns to -1 at the very end.
			
			\item 
			When we investigated the model's adherence to the null energy condition in Fig.\ref{fig:NEC}, we observed that it is violated at the same redshift that the deceleration parameter initially passes the -1 value and the Hubble parameter begins to rise. The NEC's violation suggests that dark energy density is increasing in scale factor powers, which could make the universe unstable. The statefinder analysis for the model is then performed. The anomaly is also evident here. In contrast to the standard model, which has values of  $\{r,s\}=\{1,0\}$ throughout the evolution, the PEDE model has $r$ values that are more than 1 and $s$ values that are less than zero during the late phase except the very end. Statefinder analysis reassures the presence of phantom fluid. 
			
			\item We analyzed the thermodynamic evolution of the model.
			The model doesn't meet the entropy criteria according to the generalized $2^\text{nd}$ law pertaining to the entropy of the universe. After a redshift of $z\approx -0.3$, the entropy is observed to be declining, and its first order derivative turns negative. The model also fails to meet the convexity requirement since the second derivative approaches zero from above, indicating that the entropy is not maximizing revealing an unstable asymptotic future epoch.
			
			\item We performed the phase space analysis to extract the stability conditions of the model.
			The matter-dominated epoch appeared to be unstable, as both eigenvalues were positive, implying that the universe would expand further.  However, it is found that one of the eigenvalues for the asymptotic de Sitter epoch is zero and the other is negative. The paths appeared to be converging on a line as shown in Fig.\ref{fig:phase future}. We draw the conclusion that the endpoint may be unstable based on our findings of thermodynamic analysis. The stability analysis using dark energy sound speed supports the prediction of an unstable future epoch.
			
		\end{itemize}
		
		At the outset, our general conclusion is that, even though the PEDE model may alleviate the Hubble tension, its general behavior is not all promising. The phantom behavior of the model leads to NEC violation and thermodynamically unstable future epoch. 
		
		At this juncture, we would like to mention a different phenomenological model, which is different from the present case. This model consists of two dark energy components, one of which is Einstein's cosmological constant and the other a variable dark energy component that would have emerged from the remnants of the scalar field responsible for inflation after the majority of the scalar field had transformed into matter\citep{bisnovatyi2020phenomenological,bisnovatyi2021cosmological, bisnovatyi2023eliminating}. The authors claim that the model can alleviate the Hubble tension by the action of the additional dark energy component at the stages after recombination. In the presence of the variable dark energy component, the Hubble constant decreases with time more slowly than in the $\Lambda$CDM model. The model doesn't leads to  any phantom behaviour and doesn't seem to violate the entropy maximization in the asymptotic future epoch. However, the effect of this model on CMB power spectrum and on other datasets need to be studied in detail. Within the realm of dynamical dark energy proposals, the model seems to provide a plausible explanation for the phenomenological evolution of dark energy content of the universe.
		
		\section*{Acknowledgments}
		
		One of the authors Sarath N is thankful to UGC, Govt. of India, for providing financial support through Junior Research fellowship.  The authors thank Dheepika M. and Manosh T. M. for discussions.
		
		\section*{Data Availability}
		
		The pantheon data underlying this article are available at \url{https://github.com/dscolnic/Pantheon}. The H(z) data is from Ref.\citep{Geng_2018}. The CMB data is available in Ref.\citep{chen2019distance} and BAO data is from Ref.\citep{10.1111/j.1365-2966.2011.19592.x}

		\bibliographystyle{apsrev4-2}

	\end{document}